\documentclass[conference]{IEEEtran}
\usepackage[backend=biber]{biblatex}
\usepackage[compact]{titlesec}
\usepackage{booktabs}
\usepackage[most]{tcolorbox}
\usepackage{balance}
\usepackage{flushend}

\title{Usage Patterns of Programming Languages and Tools: A Systematic Literature Review}

\author{\IEEEauthorblockN
{Sherlock A. Licorish}
\IEEEauthorblockA{\textit{University of Otago} \\
Dunedin, New Zealand \\
sherlock.licorish@otago.ac.nz \vspace{-1em}}

\and

\IEEEauthorblockN{Natalie Grattan}
\IEEEauthorblockA{\textit{University of Otago} \\
Dunedin, New Zealand \\
natalie.grattan@otago.ac.nz \vspace{-1em}}
}

\begin{document}

\maketitle

\begin{abstract}
Technology and the popularity of languages and tools are rapidly evolving. Thus, it is important to constantly assess the most popular languages and tools used in industry, furthering the collaboration between research and the software industry. In this study, we use a systematic literature review and investigate 37 studies published between 2015 -- 2024, which explored the programming languages, IDEs and CI tools used in industry. We observed that Java is most often listed among the most popular languages, and there are limited studies that investigate specific IDE usage. Also, Travis CI and GitHub Actions are the most common CI tools identified as being top CI tools used in industry. Overall, evidence shows that there is a lack of research focused on popular languages and tools used in industry, largely as there are many limitations with the possible methodology approaches. For example, lack of access to closed-source projects and sampling biases are potential issues that limit our understanding of popular languages and tools used in industry. In this regard, there is a need for more recent studies into development tools used in industry, involving multiple research approaches, to balance the limitations of each approach.
\end{abstract}


\section{Introduction}

In software engineering research, many state-of-the-art tools are developed but they are not necessarily adopted by industry. Algorithms and tools are developed by academic studies without identifying a need for such tools. The global software industry and software engineering academia have at times different priorities \cite{chimalakonda2015moving, 7890614}. In academia, they are interested in novelty and creating unique, better-performing and new tools. In industry, it is perceived, they are looking for tools that work well, not necessarily the newest tools \cite{marijan2021industry}.

With improved collaboration between industry and academia there is potential for increased insights, however, there are challenges to achieving successful collaboration \cite{barroca2018bridging}. For example, the time investment that research can involve is a barrier to industry practitioners. Software practitioners will prioritise project delivery giving pressing deadlines and may be hesitant to participate in research with a significant time commitment requirement \cite{marijan2021industry}. On the contrary, academic researchers are employed to pursue scientific inquiries. As there are challenges to academia-industry collaboration, there is significant value in summarising current knowledge of industry in research, so we can use limited resources most efficiently. 

One such knowledge area is languages and tools used in the software engineering community. Software engineering and technology move quickly as an industry. It is perceived as a highly competitive industry, with pressure put on developers to deliver software and new features rapidly \cite{rubin2016challenges}. New industry challenges such as the increased usage and developments in artificial intelligence and machine-learning, such as large-language models \cite{arora2024optimizinglargelanguagemodel}, may increase the competitive drive of the industry. As the industry moves quickly, as will the tools and programming languages used, so frequent summaries and reviews are required for academia to remain knowledgeable on current industry practices. 

It is important to know what tools and languages are being adopted by industry currently when they assemble to develop software. In order for research to be valuable to industry professionals, it should focus on what is relevant to industry. For example, if we want to create a new tool, it is more likely to be used if it prioritises languages and existing tools that are both familiar and will complement what industry professionals are already using. While the tools and languages that are being used can be found in the literature, this knowledge is scattered across multiple sources and research. 

Therefore, the goal of the current research is to systematically identify what tools and programming languages are being adopted currently in the global software industry. To achieve this goal, we investigate: (i) the programming languages, IDEs and CI tools that are reported to be used in industry; (ii) methods used to study programming languages, IDEs and CI tools; (iii) what is studied about programming languages, IDEs and CI tools; (iv) what gaps or limitations are identified. 

This work presents a Systematic Literature Review (SLR) of existing academic research focused on the tools and programming languages that are used in industry in contributing insights for the community. Given the goal of our SLR, we provide several contributions, including: (i) the programming languages, IDEs (Integrated Development Environments) and CI tools (Continuous Integration) that are reported to be used in industry, (ii) the methods used to study these tools, (iii) what is studied about the tools, and (iv) the gaps or limitations that researchers have identified. 

We found Java was the programming language most frequently identified as being a top programming language (depending on the unit of measurement each study was using). We did not find enough papers discussing IDE tools to identify the most popular IDE tools. For CI tools, Travis CI was most popular, was followed by Github Actions. The most common method of investigating languages and tools was the use of \textit{open-source} (OSS) repositories. Along with what tools and languages are used and their usage amount, factors influencing the selection of certain tools and languages over others was investigated. Researchers identified \textit{limited sample demographic} as the most common theme of study limitations. 

The remainder of the paper is organised as follows: In section \ref{section:background}, we outline relevant background research, in section \ref{section:methodology}, we outline the methodology used in the study to find the results, in section \ref{section:results}, we present the results to our four research questions, in section \ref{section:discussion}, we summarise our results and present the study implications. Finally, in section \ref{section:conclusion}, we provide concluding remarks. 

\section{Background}
\label{section:background}

In this section, we outline relevant background research, to provide context for the rest of the study.

\subsection{Continuous Integration}

\textit{Continuous integration} (CI) is a practice in software development, which usually involves automating steps such as compiling, testing and building software packages for deployment. The purpose of CI is to provide efficient feedback to developers and help meet goals of shorter release cycles \cite{fitzgerald2017continuous}. While CI is related to \textit{continuous deployment} and \textit{continuous delivery}, there are differences. Continuous deployment is considered a step further from continuous delivery \cite{shahin2017continuous}. Continuous delivery is the continuous release of software to a production environment, then continuous deployment is the continuous release of software ready to be released to users \cite{fitzgerald2017continuous}. In this SLR, the focus is on CI tools (and programming languages and IDEs), as we focus on the development environment and tools, rather than the deployment environment. 

\subsection{Related Work}

In this section, we discuss past systematic literature reviews and mapping studies that are relevant to our study. There are limited mapping studies or systematic literature reviews, which investigate what software tools and languages are used in the global software industry. The area of tools and languages taught in education and academia tended to be a more popular area of research in SLR studies \cite{noone2018visual, luxton2018introductory, medeiros2018systematic}. In Table \ref{table:past_slr} we present a summary of relevant past systematic literature reviews and mapping studies. 

Portillo-Rodríguez \textit{et al.} \cite{portillo2012tools} perform a systematic mapping study focusing on tools used for global software development. They investigate various categories of features in the tools (for example, tools that are involved in knowledge management). While this review provides a useful overview of various tools in the development process, they do not investigate programming languages or IDEs used by developers, or have a specific focus on CI tools. The study is also limited to papers prior to 2010, and as technology is a rapidly changing industry, our SLR focuses on papers published after 2010. More recently, García-Borgoñón \textit{et al.} \cite{garcia2014software} published a SLR focusing on software modelling languages used in the industry (for example, FlexUML), however, they do not focus on programming languages. 

There are more SLR studies, with a focus specifically on CI tools. Shahin \textit{et al.} \cite{shahin2017continuous} study the various tools being used in the continuous integration cycle. They find 30 methods and tools used for continuous practices, along with factors organisations need to consider when complementing software development with continuous integration, delivery and deployment tools. Similar to our SLR, Shahin \textit{et al.} \cite{shahin2017continuous} considers a range of CI tools (and continuous deployment and delivery tools). However, the focus is not on the popularity of these tools, but rather the range that is used. 

\begin{center}
\begin{table*}
\caption{Summary of Previous Software Language and Tools Systematic Literature Reviews}
\label{table:past_slr}
\centering
\begin{tabular}{p{4cm}p{2cm}p{6cm}}
\toprule
 \textbf{Authors} & \textbf{Coverage} &\textbf{Focus}\\ 
 \midrule
 Portillo-Rodríguez \textit{et al.} \cite{portillo2012tools} & 2000 -- 2010 \newline \textit{66 studies} & 132 communication, coordination and control tools for global development\\  
 García-Borgoñón \textit{et al.} \cite{garcia2014software} & 2000 -- 2011 \newline \textit{46 studies} & 40 software process modelling languages\\
 Shahin \textit{et al.} \cite{shahin2017continuous} & 2004 -- 2016 \newline \textit{69 studies} & 30 continuous practice methods and tools\\
 \bottomrule
\end{tabular}
\end{table*}
\end{center}

\section{Methodology}
\label{section:methodology}

The main goal of this systematic mapping study is to identify what tools are being used in the global software industry. Therefore, we can pave the way for future academic research, which focuses on or toward tools there is a need for. Our research methodology is based on the Kitchenham and Charter Systematic Literature Review guidelines \cite{kitchenham2007guidelines}. 

\subsection{Research Questions}

To achieve the goal of identifying what tools are currently being adopted in the global software industry, we propose the following research questions:

\subsubsection{RQ1: What programming languages, IDEs (Integrated Development Environment) and CI (Continuous Integration) tools are reported to be used in industry?}

The motivation for RQ1 is that for both a research and industry professional audience, it is useful to keep updated with what tools are being used in industry. Remaining updated means software professionals can consider using new tools that might be better suited to tasks than what they are using currently and those designing new tools know what existing tools they need to complement or integrate with. As technology is a rapidly moving industry, it is important to frequently update and revise research into what tools are being used. 

\subsubsection{RQ2: What methods are used for the study of these tools?}

The motivation of RQ2 is to investigate how researchers are currently studying tools in use. Once we know how the tools are studied, we can identify if there are ways to gain more information by using a different approach. We can also provide a reference for future researchers, who want to find the optimal approach for investigating software tools. 

\subsubsection{RQ3: What is studied about the tools?}

The motivation of RQ3 is to find what information we, as a research community, know about software tools in current use. Knowing the information that is known currently will allow for identification of gaps in knowledge. 

\subsubsection{RQ4: What gaps/limitations are identified?}

The motivation of RQ4 is to understand what is missing from current research, therefore we know what research we need to prioritise and invest resources, in future. 

\subsection{Search Strategy and Selection of Search Engines}

For the search, we intended to use a range of different search engines. Using recent software engineering systematic literature review studies as a guide \cite{wimalasooriya2022systematic, matloob2021software, NDUKWE2023111524}, we selected the following search engines: \textit{Scopus}, \textit{IEEE Xplore}, and \textit{ACM}. The goal was to ensure we searched as many sources as possible, to find a wide range of papers. Scopus, in particular, was selected as it was recommended in the Kitchenham and Charter Systematic Literature Review guidelines \cite{kitchenham2007guidelines}. We were unable to use ScienceDirect, due to limits on the number of boolean operators allowed per field and our search string being incompatible with this requirement.

As we had only found a small selection of relevant papers, we added \textit{Google Scholar}, to increase the number of possible papers that could be included in our SLR. While Google Scholar has some limitations, including lack of quality control, it is being constantly updated \cite{halevi2017suitability}. Therefore, we expected Google Scholar may have papers included that were not yet listed in the other databases. However, due to the quality limitations, we used a statistically significant sample of the estimated number of papers returned.

We also limited the search to abstract in the search engines where this capability was available (ACM, IEEE and Scopus). Further, we focused on ``computer science'' in the search engines where limitation by subject area was available (Scopus). 

\subsection{Search Strings}

To design the search string, various iterations of strings were tested and the specificity of keywords was increased until more relevant papers were returned. We aimed to achieve a balance between ensuring all relevant papers were returned, while decreasing the number of irrelevant papers. 

The first string used was: \textit{(``software industry'' OR ``developer*'' OR ``software practitioner*'') AND (``most common'' OR ``popular*'' OR ``trends'' OR ``more likely'') AND (``languages'' OR ``continuous integration techniques'' OR ``continuous integration tools'' OR ``CI techniques'' OR ``CI tools'' OR ``IDEs'' OR ``Integrated Development Environments'' OR ``questionnaire'')}. However, it was identified that some studies used the term ``CI/CD'' when discussing continuous integration tools. Additionally, most studies begin using the CI acronym, so we were able to remove some keywords to decrease the complexity of the search strings. 

The second and final search string used was: \textit{(``software industry'' OR ``developer*'' OR ``software practitioner*'' OR ``software building'') AND (``most common'' OR ``popular*'' OR ``trend*'' OR ``more likely'' OR ``numerous'' OR ``most prominent'') AND (``programming languages'' OR ``CI/CD tool*'' OR ``CI/CD service*'' OR ``CI tool*'' OR ``CI service*'' OR ``IDEs'' OR ``Integrated Development Environments'' OR ``tools and software development'' OR ``Stack Overflow'')}

\subsection{Paper Selection}

In Table \ref{table:papersreturned}, we present the number of papers returned by each search engine. The citations for Google Scholar were manually entered in a spreadsheet, as we were only using the first 30 pages. For IEEE and Scopus search engines, we could directly export the citations as a CSV file and edit them in the spreadsheet. However, for ACM, we needed to use a Python script to convert the citations to a CSV file. 

The scope of the study is limited to 2015 -- 2024. The date range was applied as a filter in all search engines.  

For Google Scholar, due to the large amount of papers returned (and as we were unable to limit to abstract, as we were with Scopus, IEEE and ACM), including less relevant papers, we took a statistically significant sample from the estimated total size provided by Google. As Google orders papers based on relevancy, we used the first 376 papers. \footnote{https://www.surveysystem.com/sscalc.htm}

For IEEE eXplore, we used the command search function and limited the search to abstracts. The search was limited to abstracts, to avoid including irrelevant papers. We also applied the date filters and conference and journal, after the search. 

For ACM, we used the \textit{ACM Guide to Computing Literature}, which is a more comprehensive database than the standard \textit{ACM Full-Text collection}. We limited to search to within abstracts, for consistency across search engines. 

For Scopus, we both limited the search to only Computer Science and to searching within the abstract. We used the advanced search functionality, to be able to apply more boolean operators to the search string. 

\begin{center}
\begin{table*}
\caption{Number of Papers Returned By Search Engine}
\label{table:papersreturned}
\centering
\begin{tabular}{p{10cm}p{3cm}}
\toprule
 \textbf{Database} & \textbf{Papers}\\ 
 \midrule
 ACM Digital Library & 62\\  
 Scopus & 754\\  
 Google Scholar & 376 \textit{(18,400)}\\
IEEE eXplore & 920\\ 
 \midrule
 \textbf{Total number of papers} & 2112\\
 \midrule
 \textbf{Total number after removing duplicates} & 1701\\
 \midrule
 \textbf{Total number after exclusion/inclusion criteria applied} & 53\\
  \midrule
   \textbf{Total number after second exclusion/inclusion round} & 37\\
  \midrule
 \textbf{Papers remaining} & 37\\
 \bottomrule
\end{tabular}
\end{table*}
\end{center}

\subsection{Inclusion and Exclusion Criteria}
\label{subsection:inclusionexclusioncriteria}

To decide if we were going to include/exclude a paper, we used the criteria outlined in Table \ref{table:criteria}. To apply the criteria, we first checked the title of the paper. If the title seemed relevant, we then checked the abstract. If from the abstract we were not able to ascertain if the paper was relevant, we then checked the research questions. 

  \begin{table*}
  \caption{Inclusion and Exclusion Criteria}
  \centering
  \label{table:criteria}
  \begin{tabular}{p{13cm}}
  \toprule
  \textbf{Inclusion Criteria}\\ 
  \midrule
   The paper is written in English.\\ 
   The paper was published in a peer-reviewed journal or conference.\\ 
   The paper is freely available to evaluate.\\
   The paper is not a secondary paper or systematic literature review or mapping study.\\
   The paper is published after 2015 and before 2025.\\
  \end{tabular}
  \begin{tabular}{ p{13cm} }
  \midrule
  \textbf{Exclusion Criteria}\\ 
  \midrule
   Papers which are a summary of a conference/workshop, thesis, editorials, keynote.\\
   Papers which do not discuss programming languages, IDEs or CI tools.\\
   Papers which infer popularity, solely based off grey literature.\\
   Papers do not focus on languages or tools in an education context.\\
   Papers that only focus on mobile development or multi-language development.\\
   Papers which specifically focus on a single technology.\\
   \bottomrule
  \end{tabular}
\end{table*}

\subsection{Data Extraction}

To extract the data, we used a spreadsheet of data fields. The extraction was a manual process undertaken by one researcher. The reason only one researcher was used for the data extraction process was so that there was consistency in the information being collected from all papers. This process involved reading each paper for the required information, then entering this information in the spreadsheet. Reliability checks were then performed by the second researcher on 10 papers, including for studies included, excluded and based on the data extracted to answer each RQ, where there was convergence.  We have included a table (Table \ref{table:data_fields}), presenting the data fields. 

\begin{center}
\begin{table*}
\caption{Data Fields Collected From Papers}
\label{table:data_fields}
\centering
\begin{tabular}{p{3cm}p{5cm}p{2cm}p{3cm}}
\toprule
 \textbf{Data Item} & \textbf{Description} & \textbf{RQ} & \textbf{Unit} \\ 
 \midrule
 Year of Publication & What is the year of publication? & Demographic Information & Year\\
 Venue & What is the name of the conference or journal the paper was published in? & & Name\\
 Conference or Journal & Is the venue a conference or journal? & & Conference or journal\\
 Country & What country are the authors from? & & Country name\\
 Affiliation & Are the authors affiliated with a university or company? & & University or company\\
 Programming languages & What programming languages are reported to be used in industry? & RQ1 & Programming languages e.g., Java, Python or C\\
 IDEs & What IDEs are reported to be used in industry? & & IDEs e.g., Netbeans or Eclipse\\
 CI Tools & What CI tools are reported to be used in industry? & & CI tools e.g., Travis CI \\
 Methods to Study Tools & What methods are used to study languages/tools? & RQ2 & Methodology e.g., survey\\
 What is Studied & What RQs are used to study languages/tools? & RQ3 & RQ types\\
 Gaps or Limitations & What gaps/limitations are there? & RQ4 & Gaps e.g., limited sample size\\
 \bottomrule
\end{tabular}
\end{table*}
\end{center}

\subsection{Quality}

As there was only a small sample of papers identified (37 papers), we did not exclude any papers based on a quality measurement. We did only consider papers that were published in peer-reviewed journals and conferences however, as these papers are required to undergo quality checks before being published. 

\section{Results}
\label{section:results}

In this section, we present the results, grouped under each research question. 

\subsection{Demographic Summary of Papers}

\begin{figure*}
  \centering
  \includegraphics[width=13cm, height=6cm]{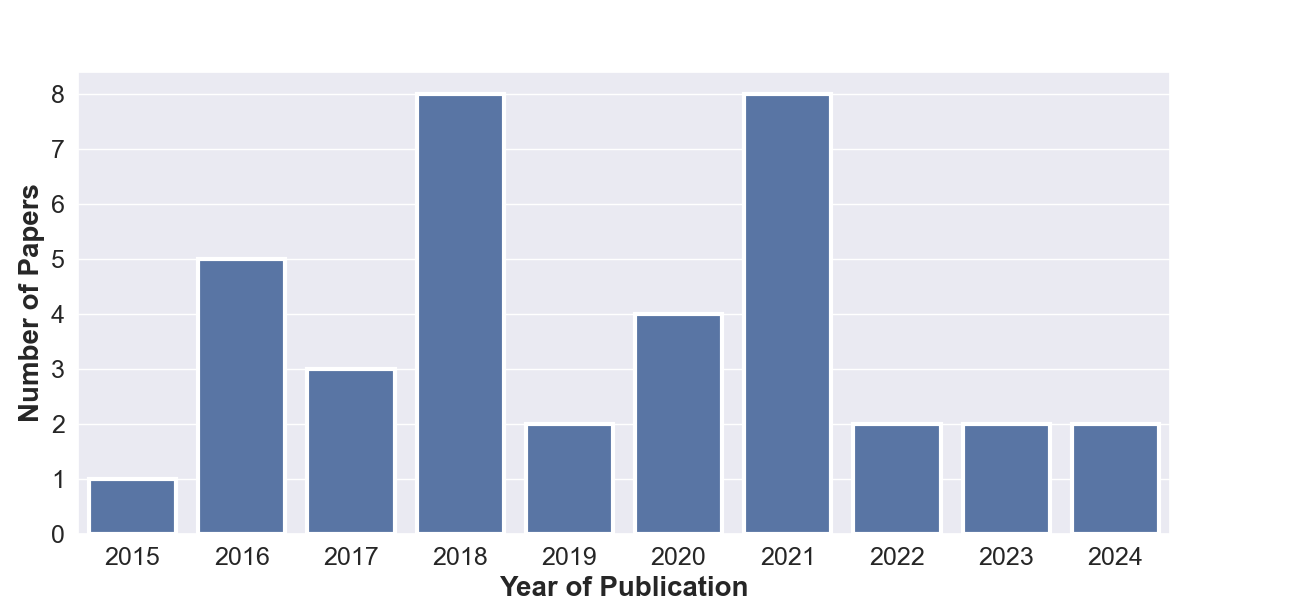}
  \caption{Number of Papers by Year of Publication}
  \label{fig:paper_years}
\end{figure*}

In this section, we present demographic information about the papers selected for our study. We have the complete list of papers in the appendices. The total number of studied papers was 37. We present the number of papers listed by the year of publication in Figure \ref{fig:paper_years}. We see the years with the most papers published are 2018 and 2021 (8 papers in each year). The number of papers decreases significantly between 2021 and 2022, then stabilises. 

\begin{figure*}
  \centering
  \includegraphics[width=13cm, height=6cm]{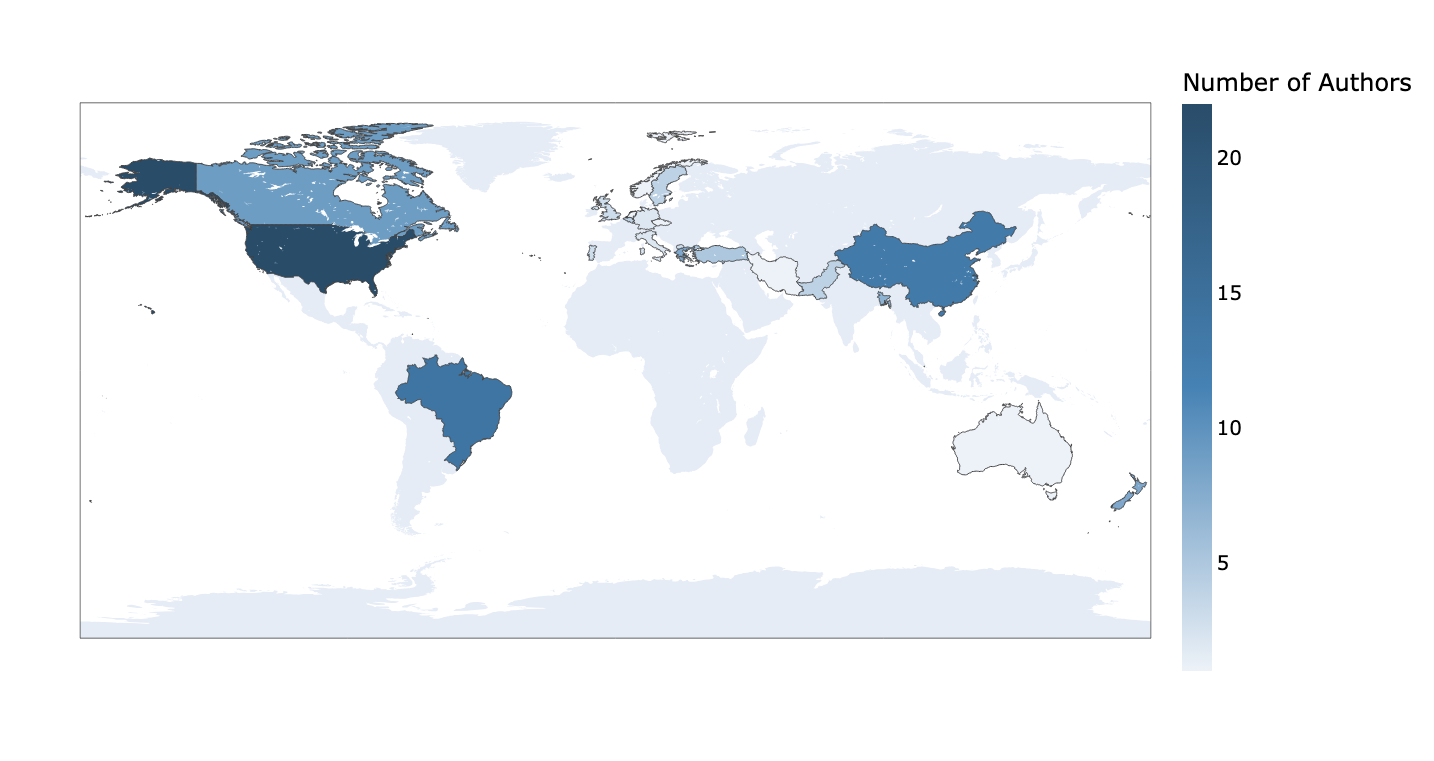}
  \caption{Number of Authors by Industry/University Country}
  \label{fig:authors_countries}
\end{figure*}

Figure \ref{fig:authors_countries} presents the number of authors affiliated with universities and organisations in each country. From the data shown in Figure \ref{fig:authors_countries}, it is clear the majority of authors are affiliated with North American universities and organisations (22 authors). This is expected, as the software industry in the USA is very large \cite{arora2005globalization}. There were also high numbers of authors from Canada, Brazil, and China. However, we had no authors from Indian universities and organisations, where there are also large software industries \cite{arora2005globalization}, which was a more unexpected result. 

\subsection{RQ1: What programming languages, IDEs (Integrated Development Environment) and CI (Continuous Integration) tools are reported to be used in industry?}

\begin{tcolorbox}[enhanced,attach boxed title to top center={yshift=-3mm,yshifttext=-1mm},
  colback=gray!5!white,colframe=gray!75!black,colbacktitle=gray!80!black,
  title=RQ1: Summary,fonttitle=\bfseries,
  boxed title style={size=small,colframe=black}]
We find the most popular programming languages to be Javascript, Java, and Python. Very few papers focused on IDEs. The most popular CI service was Travis CI.
\end{tcolorbox}

\subsubsection{Programming Languages}

As each study highlights a different number of languages, we have generally limited this section to the \textit{top-3 languages} discussed in each study. In some papers, there is a fourth or fifth language discussed as it is equal to one of the top-3 languages. We have also grouped together C and C++, the .NET programming languages (C\#, F\# and Visual Basic), Python and Ruby, HTML and CSS, and PHP and Perl. These languages were often grouped together in the respective studies. As each study has a different criteria for measuring what a top language is, we use the languages identified by the study. 

\begin{figure*}
  \centering
  \includegraphics[width=13cm, height=6cm]{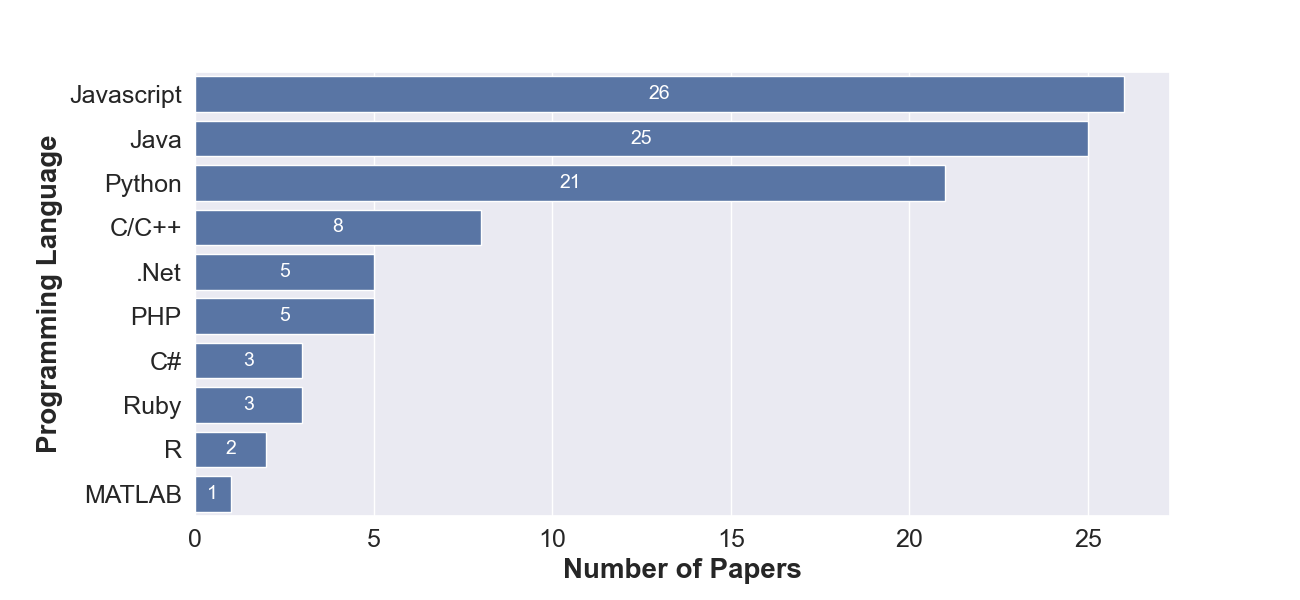}
  \caption{Top Languages in the Papers Studied}
  \label{fig:top_languages}
\end{figure*}

In Figure \ref{fig:top_languages}, we present the top languages in the papers studied. The language that was identified in the top languages most often was Java, followed by Javascript and Python. In comparison to the TIOBE index (as at July 2024) \cite{tiobe2}, Java is fourth and Python is first, so these results were slightly unexpected. A possible reason for the higher ranking of Python on the TIOBE index, which is more updated than the studies in this SLR, is due to recent \textit{large language model} (LLM) popularity and Python tends to be a more popular language in data science and machine learning. 

\subsubsection{IDEs}

Overall, we only identified one study which investigated IDE usage. Therefore, we were unable to answer the IDE part of RQ1. However, despite not being able to identify the popular IDEs used in the software industry, we have learned studying IDEs is a potential area of study for future research. It is possible there is less interest in IDEs, as it could be more difficult to collect data due to the IDE being used on local computers and being more difficult to ascertain from open-source software projects. However, while popular languages and CI tools could be developers using what they are told to use in their job, IDE usage could be a more personal decision and therefore an interesting area of research.

\subsubsection{CI Tools}

As with studies focusing on IDEs, there were less studies with a focus on the use of specific CI tools in comparison with studies focusing on the use of specific programming languages (P4, P14, P19, P25, P31, P32). 

\begin{figure*}
  \centering
  \includegraphics[width=13cm, height=6cm]{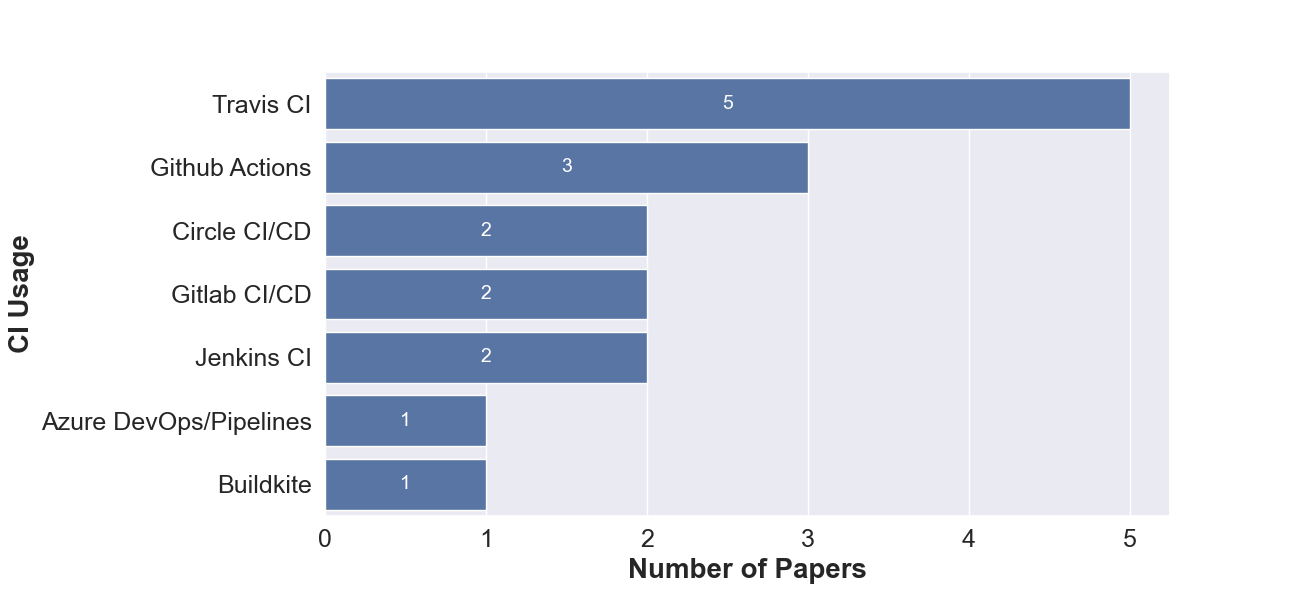}
  \caption{Top CI Tools Used in the Papers Studied}
  \label{fig:top_ci}
\end{figure*}

In Figure \ref{fig:top_ci} we present the top CI tools used by or in demand by developers/industry professionals, found by the papers studied in this SLR. We find that Travis CI is the most common tool identified by researchers as being a top tool used, followed by GitHub Actions. 

\subsection{RQ2: What methods are used for the study of these tools?}

\begin{tcolorbox}[enhanced,attach boxed title to top center={yshift=-3mm,yshifttext=-1mm},
  colback=gray!5!white,colframe=gray!75!black,colbacktitle=gray!80!black,
  title=RQ2: Summary,fonttitle=\bfseries,
  boxed title style={size=small,colframe=red!50!black} ]
  The most common way of studying software tools is by mining open-source repositories, such as Github, followed by using Stack Overflow.
\end{tcolorbox}

In order to present a summary of the results, we generalised the method types using categories. These were generated using a bottom-up analysis technique, by which we took the specific method from each study, then we manually created labels based on commonalities between the methods. Some studies used more than one method (P1, P4, P11, P12, P16, P20, P28, P34) to study the tools. 

\begin{figure*}
  \centering
  \includegraphics[width=13cm, height=6cm]{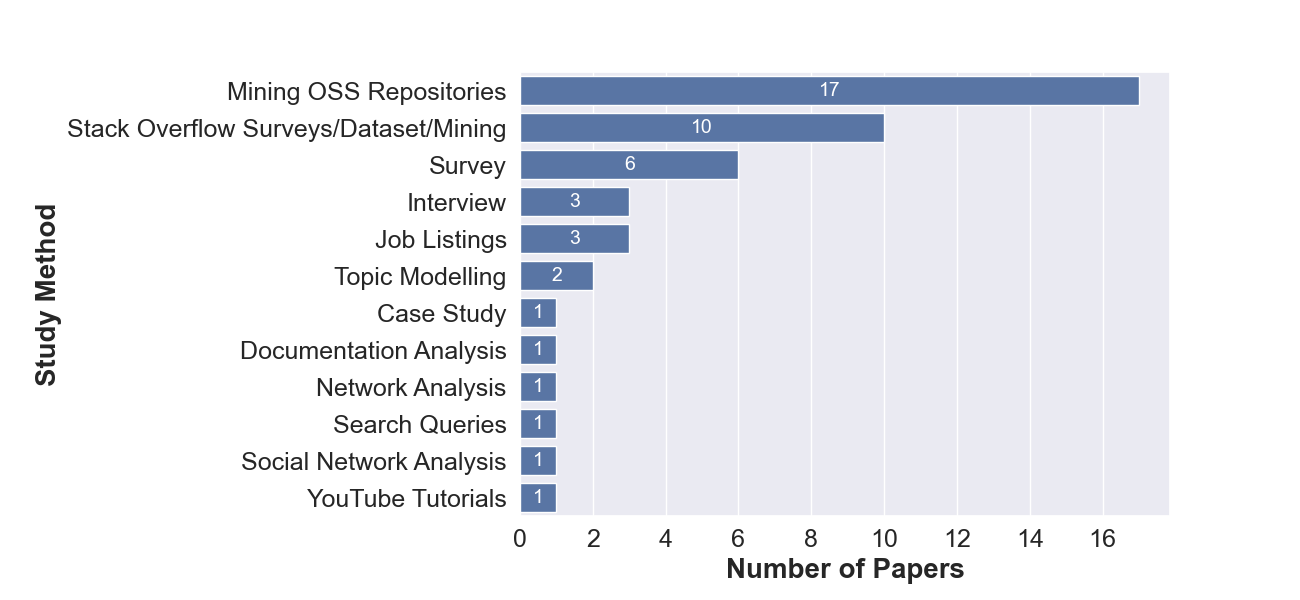}
  \caption{Methods Used to Study Languages/Tools in the Papers Studied}
  \label{fig:methods}
\end{figure*}

In Figure \ref{fig:methods} we present the method types used to study the programming languages and tools. The most common way is by mining open source data repositories (generally using the Github API). The second most common approach was using Stack Overflow, generally by mining different questions and answers. 

\subsection{RQ3: What is studied about the tools?}

\begin{tcolorbox}[enhanced,attach boxed title to top center={yshift=-3mm,yshifttext=-1mm},
  colback=gray!5!white,colframe=gray!75!black,colbacktitle=gray!80!black,
  title=RQ3: Summary,fonttitle=\bfseries,
  boxed title style={size=small,colframe=red!50!black} ]
  The main themes studied about the tools are identifying what tools/languages are used and the usage amount (which was measured in various different ways).
\end{tcolorbox}

We apply a thematic analysis to investigate what different papers are studying about programming languages and tools. For RQ3, we studied both the research questions, which relate to the tools and languages and the study results. In order to apply the thematic analysis, we used an iterative process. We looked at what was studied about the languages and tools for each paper and found common themes. We then studied each of the themes found to see if any could be combined or should be expanded on. 

\begin{figure*}
  \centering
  \includegraphics[width=13cm, height=6cm]{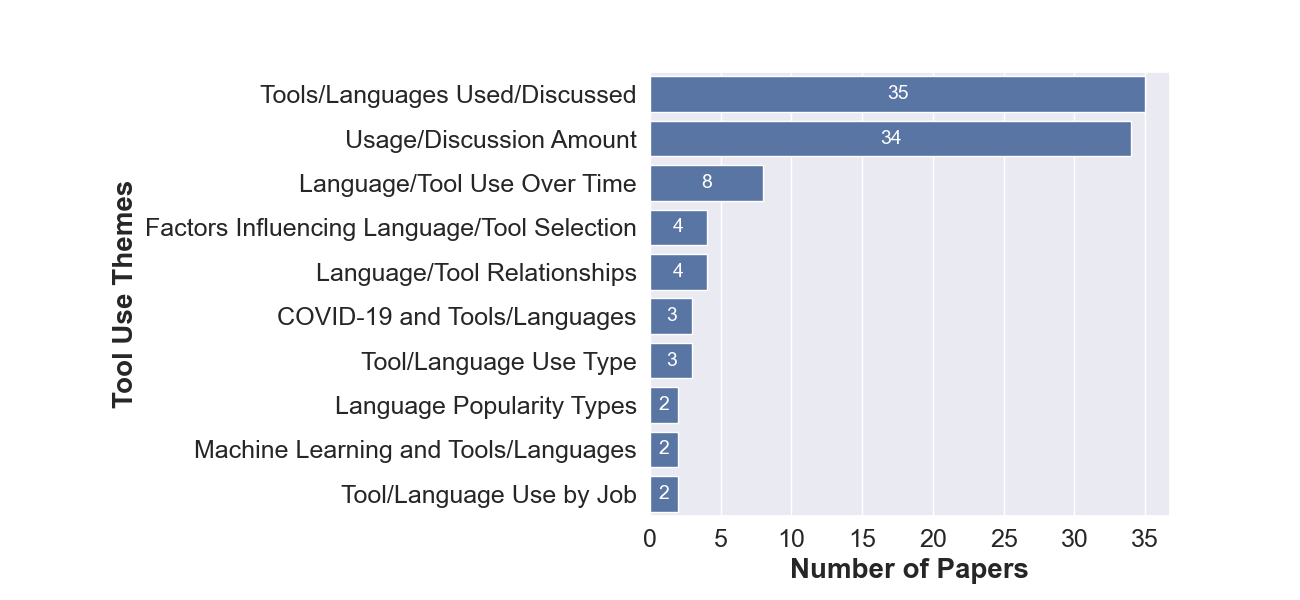}
  \caption{Top Themes Representing What is Studied About Tools/Languages}
  \label{fig:use_themes}
\end{figure*}

In Figure \ref{fig:use_themes} we present the most common themes (all the themes present in more than one paper), representing what researchers study about programming languages and software tools. By far, the most common two themes were \textit{types of language/tool use} and \textit{usage/discussion amount}. The reason these themes are the most common is likely because these tend to form a baseline and lead into further discussing and analysing of tools and languages. This tends to be in the form of a table, figure or explanation of what tools/languages are used and the amount presented. Therefore, we always see these themes occurring together. 

The next most common themes are \textit{language/tool use over time}, \textit{factors influencing language/tool selection}, and \textit{language/tool relationships}. Language/tool use over time focuses on the trends, such as if certain tools or languages are being used more frequently today than in previous years. For example, in P15, the authors use a line chart to show the popularity of languages in Stack Overflow posts between 2008 -- 2020. Factors influencing language/tool selection involves when the authors discuss what features influence the use (or lack of use) of particular languages or tools. For example, in P25, they discuss the reasons reported for using particular CI tools, through an interview/questionnaire, such as ``increased reliability'' or ``increased quality''. Language/tool relationships goes beyond just factors, to whether the authors discuss correlations. For example, in P31, they find popular projects are more likely to use CI tools. 

\subsection{RQ4: What gaps/limitations are identified?}

\begin{tcolorbox}[enhanced,attach boxed title to top center={yshift=-3mm,yshifttext=-1mm},
  colback=gray!5!white,colframe=gray!75!black,colbacktitle=gray!80!black,
  title=RQ4: Summary,fonttitle=\bfseries,
  boxed title style={size=small,colframe=red!50!black} ]
  The most frequent limitations identified by study authors were limited sample demographic and limited generalisability.
\end{tcolorbox}

We use a thematic analysis to find any gaps/limitations identified by the authors. In some cases, the authors did not identify any gaps or limitations (P7, P10, P12, P20, P21, P23, P24, P36, P37), however that does not necessarily signify there were no limitations in the study method, rather that the researchers did not have space due to word limitations of particular venues or none of the threats were perceived salient enough to comment on. These gaps/limitations are limitations identified by the researchers, rather than the authors of this SLR. 

\begin{figure*}
  \centering
  \includegraphics[width=13cm, height=6cm]{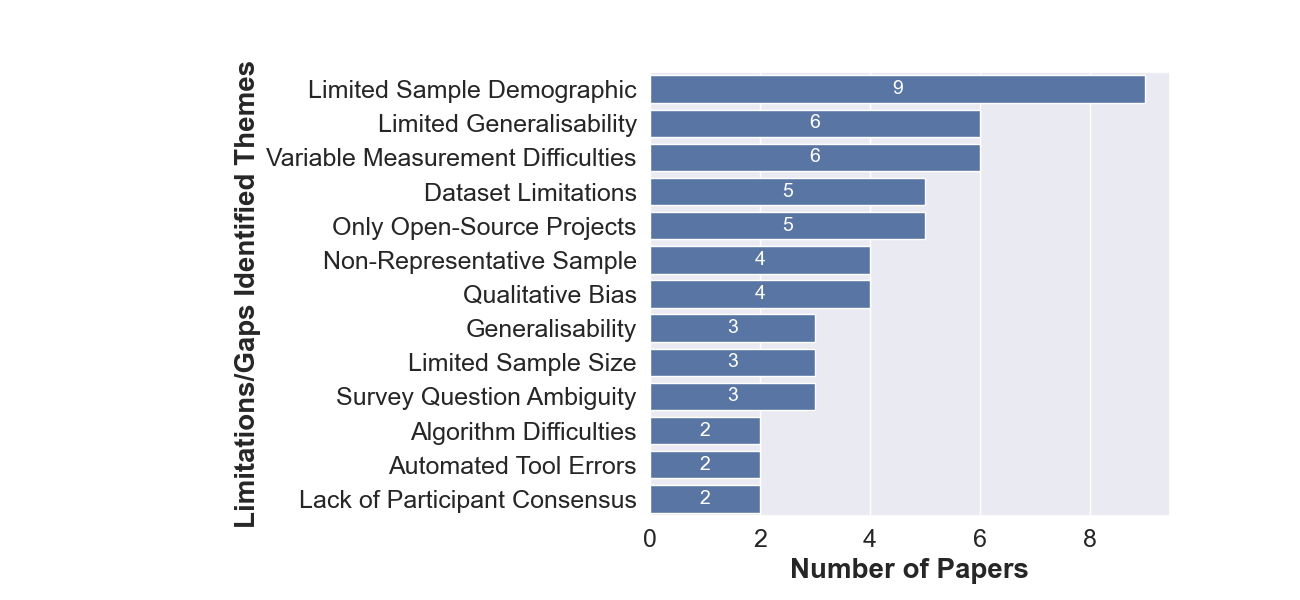}
  \caption{Top Themes Representing Gaps/Limitations Researchers Identified}
  \label{fig:limitations_themes}
\end{figure*}

In Figure \ref{fig:limitations_themes} we present the most common themes (all the themes present in more than one paper). The most frequent theme present is \textit{limited sample demographic}, followed by \textit{limited generalisability}. Limited sample demographic is when the researchers were unable to be completely representative of the group they were investigating. Limited generalisability is related to how representative the sample was, but this label was applied when the researchers specifically identified the limitation as having limited generalisability. For example, both themes were presented in P17. There was a limited sample demographic, as they mention their study is ``skewed toward developers in Sweden and China''. There was also limited generalisability as they later mention the limited sample demographic might mean there is a threat when findings are applied to a ``broader population who do not share these characteristics''. 

\section{Implications, Threats and Future Work}
\label{section:discussion}

\subsection{Implications}

\subsubsection{Implications for Researchers}

There are many flow-on implications of this research for the wider academic community. When designing a new tool, researchers would be advised to design the tool to be compatible with the programming languages, IDEs and CI tools that are currently being used by industry developers and practitioners. Designing the tool to be compatible with the most popular development tools would make it more widely accessible and therefore, more likely to be used. Additionally, learning of different methodologies used to investigate tools and languages can help researchers identify possible approaches when wanting to answer research questions regarding tools and languages. 

\subsubsection{Implications for Practitioners}

The results of this study will also have many implications for practitioners. To summarise, it is hoped that the results will help software practitioners improve their software methodology processes. Knowing the most common tools and languages being used can help decide which tools and languages to use for upcoming projects or ensure there is compatibility with global technology. In particular, for developers designing software tools meant to be used by other developers, such as Infrastructure as a Service (IaaS) or Platform as a Service (PaaS), it is necessary for developed tools to be compatible with those that are widely used. The results also provide practitioners a more high-level view of the state of the industry, in terms of what tools are being used widely. Knowing the state of the industry can help a software organisation ensure their tech stack is up-to-date with industry standards and whether they should consider upgrading or adding compatibility with a new IDE. 

\subsection{Threats to Validity}

Validity is an assessment of how accurately the study and methods measure what they are intended to measure \cite{messick1987validity}. In SLR studies, we need to consider: descriptive validity, theoretical validity, generalizability and
interpretive validity \cite{petersen2015guidelines}. In this section, we address potential threats to validity. 

\subsubsection{Descriptive Validity}

\textit{Descriptive validity} is how accurately results have been described \cite{petersen2015guidelines} and is an important consideration in SLR studies, as they utilise qualitative research methods. As the data collection was performed by a single researcher with reliability checks done by another, there is the possibility of some biased results. To uphold descriptive validity, we used a data form (see Table \ref{table:data_fields}) to collect data from the papers considered in the study. 

\subsubsection{Theoretical Validity}

\textit{Theoretical validity} is related to ensuring our set of studies was complete and that we did not miss any relevant studies \cite{petersen2015guidelines}. To try to ensure we collected all relevant studies, we used four different search engines. 

\subsubsection{Generalisability}

\textit{Generalisability} refers to how well the results of our study can be applied in a wider context \cite{petersen2015guidelines}. Our study focuses on the general software industry, and it is important to note that the software industry is very large and it is possible some areas were missed (for example, the software industry in some countries may not be represented). However, we followed a documented search process to find the studies included and used four different search databases, to try to ensure we had the most representative sample of papers possible. 

\subsubsection{Interpretive Validity}

\textit{Interpretive validity} means how reasonable the conclusions drawn from the data are \cite{petersen2015guidelines}. As SLRs are qualitative studies and require the use of classification schemes in order to be able to make generalisations from the results, there is a possibility some researchers may disagree with how certain data points have been classified. We are  therefore willing to make the data collected publicly available. 

One of the challenges in interpretive validity is that it can be difficult to compare the most commonly used tools, as in each study, there are different conditions surrounding the use of these tools. For example, the most discussed tools and languages may not necessarily be the most used tools and languages. However, as we are focusing on studies holistically, across a range of conditions, we are able to see trends of which tools and languages are consistently reported across a variety of conditions. 

Another challenge is while the objective of this study was to focus on what programming languages and tools are being used by industry software developers, depending on what the researchers are measuring, there is the possibility researchers or students have been included. For example, if the study is using Stack Overflow to measure what programming languages are being commonly discussed, it is highly likely this sample will include researchers and students. We did exclude studies where the sample was very likely to contain only researchers or students, or it was specifically mentioned researchers and students were targeted. 

\subsection{Future Work}

As we found only a small number of studies investigating programming languages and tools, overall there is significant potential in this area, particularly for investigating what IDEs developers are using. When consulting with industry, it is important to be aware of differing priorities between researchers and industry professionals. For example, a priority in research is to ensure replicability and therefore visibility of the research methodology. However, in industry, there can be concerns of releasing proprietary information which can limit information that can be used and published. Additionally, in industry, taking time out of a busy schedule to answer surveys may not be seen as an efficient use of time, especially if there is a perception of a large gap between academia and industry. If the survey is unlikely to have any practical consequences, there can be less motivation to spend time filling it in. Therefore, longitudinal studies, which may take a significant time commitment from the software professionals may not be the best approach, if they can be avoided. 

A current survey of developers and software professionals is an approach that would provide useful information. A survey could also provide information that many studies (particularly those employing quantitative analysis tools) are missing, such as the factors influencing tool usage. However, as found in RQ4, there are many limitations associated with surveys. For example, ambiguity in survey questions (which was identified in P5, P18 and P25). Developers and software professionals are busy, so there is a balance to be achieved in ensuring a survey is clear, but if it is too long and wordy, there is a chance less developers will participate. This can result in a self-selection bias. Therefore one approach may be to combine a survey with mining software repositories or forums such as Stack Overflow, so the limitations can be balanced out. 

\section{Conclusion}
\label{section:conclusion}

In conclusion, the main goal of this research was to investigate what knowledge we have in academia about what tools and programming languages are currently being used in industry. We found the programming languages that were most commonly identified as the top languages used were Javascript, Java, and Python. However, there was not enough research to draw any conclusions on the popularity of particular IDEs. That said, the most common CI tools were Travis CI and Github Actions. The most common study methodology for tools and languages was mining OSS repositories, closely followed by using a developer survey. These correspond with the most common gaps/limitations identified by researchers, including limited sample demographic, limited generalisability and survey question ambiguity. Overall, there was limited research available in the area of industry tools and languages, particularly for IDE usage. Therefore, there is significant research potential in this area, particularly research combining multiple methodology approaches. Research should particularly look at ways to engage commercial developers about their technology use to represent knowledge of the broad software engineering community.

\printbibliography

\section*{Appendix: Selected Studies}

\begin{itemize}
  \item P1: A decision model for programming language ecosystem selection: Seven industry case studies (2021)
  \item P2: A large scale study of long-time contributor prediction for GitHub projects (2019)
  \item P3: A large scale study of multiple programming languages and code quality (2016)
  \item P4: A mixed method study of DevOps challenges (2023)
  \item P5: A survey on modeling and model-driven engineering practices in the embedded software industry (2018)
  \item P6: An analysis of 35+ million jobs of Travis CI (2019)
  \item P7: An analysis of user profiles from Covid-19 questions in Stack Overflow (2022)
  \item P8: An empirical analysis of the utilization of multiple programming languages in open source projects (2015)
  \item P9: An empirical study of COVID-19 related posts on Stack Overflow: Topics and technologies (2021)
  \item P10: An exploratory study on how software reuse is discussed in Stack Overflow (2020)
  \item P11: Analysis of software engineering industry needs and trends: Implications for education (2017)
  \item P12: Analysis of the popularity of programming languages in open source software communities (2020)
  \item P13: Can you tell me if it smells? A study on how developers discuss code smells and anti-patterns in stack overflow (2018)
  \item P14: Characterizing the usage of CI Tools in ML projects (2022)
  \item P15: Community evolution on Stack Overflow (2021)
  \item P16: Data driven development trend analysis of mainstream information technologies (2016)
  \item P17: Developers' perspective on today's and tomorrow's programming tool assistance: A survey (2024)
  \item P18: Development processes and practices in a small but growing software industry: A practitioner survey in New Zealand (2018)
  \item P19: Does using Bazel help speed up continuous integration builds? (2024)
  \item P20: Evaluating a programming topic using GitHub data: what we can learn about machine learning (2021)
  \item P21: Exploring industrial demand trend’s in Pakistan software industry using online job portal data (2017)
  \item P22: Googling for software development: What developers search for and what they find (2021)
  \item P23: Identifying trends in technologies and programming languages using topic modeling (2018)
  \item P24: Investigating trends in learning programming using YouTube tutorials (2020)
  \item P25: On the usage, co-usage and migration of CI/CD tools: A qualitative analysis (2023)
  \item P26: Oops, my tests broke the build: An explorative analysis of Travis CI with GitHub (2017)
  \item P27: Representation of developer expertise in open source software (2021)
  \item P28: Software development during COVID-19 pandemic: An analysis of Stack Overflow and GitHub (2020)
  \item P29: The state of the ML-universe: 10 Years of artificial intelligence \& Machine Learning Software Development on GitHub (2018)
  \item P30: Understanding the factors that impact the popularity of GitHub repositories (2016)
  \item P31: Usage, costs, and benefits of continuous integration in open-source projects (2016)
  \item P32: Use and misuse of continuous integration features: An empirical study of projects that (mis) use Travis CI (2018)
  \item P33: Which programming language and platform developers prefer for the development? A study using Stack Overflow (2018)
  \item P34: Why and how developers fork what from whom in GitHub (2016)
  \item P35: Work practices and challenges in continuous integration: A survey with Travis CI users (2018)
  \item P36: What programming languages do developers use? A theory of static vs dynamic language choice (2018)
  \item P37: What are the information technology skills needed in information institutions? The case of ``code4lib'' job listings (2021)
\end{itemize}

\end{document}